\documentclass[aps,twocolumn,prl,superscriptaddress]{revtex4}
\usepackage{epsfig}
\usepackage{amsmath}
\newcommand{\ben}{\begin{eqnarray}}
\newcommand{\een}{\end{eqnarray}}
\newcommand{\nnu}{\nonumber\\}
\newcommand{\bef}{\begin{figure}[htb]\centering}
\newcommand{\eef}{\end{figure}}
\newcommand{\sla}[1]{\slash\!\!\!{#1}}
\newcommand{\p}{\mathcal{P}}
\newcommand{\cp}{\mathcal{CP}}
\newcommand{\nbar}{\bar{n}}

\begin{document}
\title{Quark fragmentation in the $\theta$-vacuum}
\author{Zhong-Bo Kang}
\affiliation{RIKEN BNL Research Center,
                Brookhaven National Laboratory,
                Upton, NY 11973, USA}
                
\author{Dmitri E. Kharzeev}
\affiliation{Department of Physics, 
                Brookhaven National Laboratory,
                Upton, NY 11973, USA}
                
\date{\today}
\begin{abstract}
The vacuum of Quantum Chromodynamics is a superposition of degenerate states with different topological numbers 
that  are connected by tunneling (the $\theta$-vacuum). The tunneling events are due to 
topologically non-trivial configurations of gauge fields (e.g. the instantons) that induce local $\p$-odd domains in Minkowski space-time.
 We study the quark fragmentation in this topologically non-trivial QCD background. 
 We find that even though QCD globally conserves $\p$ and $\cp$ symmetries, two new kinds of $\p$-odd fragmentation functions emerge. They generate interesting dihadron correlations:  one is the azimuthal angle correlation $\sim \cos(\phi_1 + \phi_2)$ usually referred to as the Collins effect, and the other is the $\p$-odd correlation $\sim \sin(\phi_1 + \phi_2)$ that vanishes in the cross section summed over many events, but survives on the event-by-event basis. Using the 
chiral quark model we estimate the magnitude of these new fragmentation functions. We study their experimental manifestations in dihadron production in $e^+e^-$ collisions, and comment on the applicability of our approach in deep-inelastic scattering, proton-proton and heavy ion collisions.
\end{abstract}

\maketitle

%%%%%%%%%%%%
{\it{1. Introduction.}}
Quantum Chromodynamics (QCD) is at present firmly established as 
the theory of the strong interactions. Equations of motion in QCD possess topologically non-trivial solutions \cite{Belavin:1975fg} signaling the presence of degenerate ground states differing by the value of topological
charge \cite{Chern:1974ft}.
The physical vacuum state of the theory is a superposition of these degenerate
states, so-called $\theta$-vacuum \cite{'tHooft:1976up}. To reflect this vacuum structure one may equivalently introduce a $\theta$-term in the QCD Lagrangian. Unless $\theta$ is identically
equal to zero, this term explicitly breaks $\p$ and $\cp$ symmetries of QCD. However  
stringent limits on the value of $\theta<3\times 10^{-10}$
deduced from the experimental bounds on the electric dipole moment of the 
neutron \cite{Baker:2006ts} indicate the absence of {\it global} $\p$ and $\cp$ violation
in QCD. 

Nevertheless it has been proposed that the {\it local} $\p$- and $\cp$-odd effects due to 
the topological fluctuations characterized by an effective $\theta=\theta(\vec{x}, t)$ varying in space and time
could be directly observed through multi-particle correlations \cite{Kharzeev:1998kz}. 
In heavy ion collisions, the existence of magnetic field (and/or the angular momentum) in the presence of topological 
fluctuations can induce the separation of electric charge with respect to the reaction plane, so-called Chiral Magnetic Effect  \cite{Kharzeev:2004ey, Kharzeev:2007tn, Kharzeev:2007jp, Fukushima:2008xe}.
There is a recent experimental evidence for this effect from STAR Collaboration at RHIC \cite{star:2009uh}. 
The interpretation of STAR result in terms of the local parity violation is under intense scrutiny at present, see e.g.
 \cite{Bzdak:2009fc,Wang:2009kd, Asakawa:2010bu}.

In this paper, we study the role of QCD topology in hard processes using the 
formalism based on factorization theorems \cite{Collins:1989gx}.  From the QCD factorization point of view, the cross section
in high energy collision can be factorized into a convolution of perturbatively calculable partonic cross section and the non-perturbative but universal parton distribution and fragmentation functions. In the conventional formalism, these distribution and fragmentation functions are required to be $\p$-even because of the parity-conserving nature of the strong interaction. However, in the presence of local (in space and time) $\p$-odd domains $\p$-odd fragmentation functions can emerge \cite{Efremov:1995ff}; note that only the cross section of the entire process has to be $\p$-even, not the fragmentation function. 

In this letter, we derive the most general form of the quark fragmentation function for a quark fragmenting into a pseudoscalar meson which is consistent with the Lorentz invariance. Abandoning the parity constraint, we obtain two $\p$-odd fragmentation functions besides the 
well-known $\p$-even spin-averaged fragmentation function \cite{Collins:1981uw} and Collins function \cite{Col93}. 
We obtain the exact operator definitions and estimate the size of these new $\p$-odd fragmentation functions
using the chiral quark model \cite{Manohar:1983md}.
As a first step, we present their observable effect in the back-to-back dihadron production in $e^+e^-$ collisions. We encourage the experimentalists to carry out the related analyses at RHIC and elsewhere.

%%%%%%%%%%%%
{\it 2.~Quark fragmentation functions in locally $\p$-odd background.}
The quark fragmentation functions are defined through the following matrix \cite{Bacchetta:2006tn}:
\ben
\Delta\left(z, p_\perp\right)
&=&\frac{1}{z}\int\frac{dy^-d^2y_\perp}{(2\pi)^3}e^{ik\cdot y}
\langle 0| {\cal L}_y \psi(y)|P  X\rangle
\nnu
&&\,
\langle P  X| \bar{\psi}(0){\cal L}_0^\dagger|0\rangle|_{y^+=0},
\een
where
$p$ is the momentum of the final state hadron with a transverse momentum $p_\perp$ relative to the fragmenting quark $k$.
We choose the hadron moving along $+\hat z$ direction, and 
define the light-cone momentum $p^\pm=(p^0\pm p^z)/\sqrt{2}$. 
For convenience, we define two light-like vectors:
$\nbar^\mu=\delta^{\mu+}$ and $n^\mu=\delta^{\mu-}$.
The momentum fraction $z=p^+/k^+$, and $\vec{k}_\perp=-\vec{p}_\perp/z$.
${\cal L}_y={\cal P}\exp\left(ig\int_0^\infty d\lambda n\cdot A(y+\lambda n)\right)$ is the gauge link needed to make $\Delta\left(z, p_\perp\right)$ gauge invariant.

Since QCD is a theory conserving $\mathcal{C}$, $\mathcal{P}$, and $\mathcal{T}$ globally, one usually expands the above matrix using the following constraints \cite{Boer:2003cm}:
\ben
{\rm Hermiticity:}&&\qquad \Delta^\dag(p, k)=\gamma^0  \Delta(p, k) \gamma^0
\\
{\rm Parity:}&&\qquad \Delta(p, k)=\gamma^0  \Delta(\bar{p}, \bar{k}) \gamma^0
\label{parity}
\\
{\rm Time-reversal:}&&\qquad \Delta^*(p, k)=V_T  \Delta(\bar{p}, \bar{k})V_T^{-1}
\een
where $V_T=i\gamma^1\gamma^3$ and $\bar{p}^\mu=p_\mu=(p^0, -\vec{p})$. Using the basis of gamma matrices $\Gamma=\{1, \gamma^\mu, \gamma^\mu\gamma^5, \sigma^{\mu\nu}, i\gamma^5\}$, and the available momenta $p$ and $k$, one can expand $\Delta(p, k)$ in the most general form:
\ben
\Delta(p, k)&=&\left[M A_1 1+A_2\sla{p}+A_3\sla{k}+A_4\sigma^{\mu\nu}\frac{k_\mu p_\nu}{M}\right]
\nnu
&&+\bigg[A_5\sla{p}\gamma^5+
A_6\sla{k}\gamma^5+M A_7 i\gamma^5
\nnu
&&\left.
+A_8\sigma^{\mu\nu}i\gamma^5\frac{k_\mu p_\nu}{M}\right],
\label{deltapk}
\een
where $M$ is the hadron mass used to make all $A_i$'s have the same dimension. Since the time-reversal changes out-state to in-state, it does not really give any constraint on the coefficients $A_i$ \cite{Boer:2003cm}. One the other hand, if one applies the Hermiticity constraint, all of the $A_i$'s have to be real. If one further applies Parity constraint, one finds $A_5=A_6=A_7=A_8=0$. However, as we stated in the Introduction, 
we are interested in the situation in which a local $\p$-odd domain develops in space-time, and the quark fragmentation happens inside such a $\p$-odd domain (or in other words, the quark scatters off the non-trivial gauge field configuration prior to transforming into a pseudoscalar meson). In this case, the $\p$-odd modes in the quark fragmentation could be populated and one has to release the parity constraint in Eq.~(\ref{parity}). Without parity constraint, we thus need to keep all 8 terms $A_1$ through $A_8$ in Eq.~(\ref{deltapk}). Applying the twist-expansion by parametrizing the momenta as $p^\mu\approx p^+ \bar{n}^\mu$ and $k^\mu\approx {(p^+\bar{n}^\mu-p_\perp^\mu)}/z$ and keeping the leading terms we obtain
\ben
\Delta(z, p_\perp)&=&\frac{1}{2}\left[D(z, p_\perp^2)\sla{\nbar}+H_1^\perp(z, p_\perp^2) \sigma^{\mu\nu}\frac{p_{\perp\mu} \nbar_\nu}{M}\right]
\nnu
&&+\frac{1}{2}\bigg[ \widetilde{D}(z, p_\perp^2)\sla{\nbar}\gamma^5
\nnu
&&\left.
+\widetilde{H}_1^\perp(z, p_\perp^2) \sigma^{\mu\nu}i\gamma^5\frac{p_{\perp\mu} \nbar_\nu}{M}
\right].
\label{main}
\een
where $D(z, p_\perp^2)$  and $H_1^\perp(z, p_\perp^2)$ are the usual $\p$-even fragmentation functions: $D(z, p_\perp^2)$ is the transverse momentum dependent spin-averaged fragmentation function \cite{Collins:1981uw}, and $H_1^\perp(z, p_\perp^2)$ is the Collins function describing a transversely polarized quark fragmenting into an unpolarized hadron \cite{Col93}. % with the transverse momentum relative to the parent quark momentum correlated with the transverse polarization vector of the fragmenting quark. 
Now besides the two conventional $\p$-even fragmentation functions, we also obtain two new $\p$-odd fragmentation functions: $\widetilde{D}(z, p_\perp^2)$ and $\widetilde{H}_1^\perp(z, p_\perp^2)$. As we will show below, $\widetilde{H}_1^\perp(z, p_\perp^2)$  generates a new kind of azimuthal correlation. Its role is similar to $H_1^\perp(z, p_\perp^2)$: $H_1^\perp(z, p_\perp^2)$ represents an asymmetric distribution $\propto ({\hat p}\times p_\perp)\cdot\vec{s}_q$, while $\widetilde{H}_1^\perp(z, p_\perp^2)$ represents an asymmetric distribution $\propto p_\perp\cdot \vec{s}_q$ for a transversely polarized quark with spin vector $\vec{s}_q$ to fragment into a pseudoscalar meson. The newly derived $\p$-odd
fragmentation functions will lead to interesting $\p$-odd effects in experiment as we will show in the next section.

In order to study the experimental effects generated by these $\p$-odd fragmentation functions, we need to estimate their magnitude. For this purpose we use the effective chiral quark model developed by Manohar and Georgi \cite{Manohar:1983md}, which is an effective theory of QCD at low energy scale. This model has also been adopted for an estimate of the Collins functions in \cite{Bacchetta:2002tk, collins}.
The effective Lagrangian describing the interaction between the quarks and the pion in the leading order is given by
\ben
L_{qq\Pi}=-\frac{g_A}{2f_\pi}\bar{\psi}_q\gamma^\mu\gamma^5 \vec{\tau}\cdot\partial_\mu \vec{\pi}  \psi_q
\label{lqqpi}
\een
where $f_\pi\approx 93$ MeV is the pseudoscalar decay constant.
\bef
\psfig{file=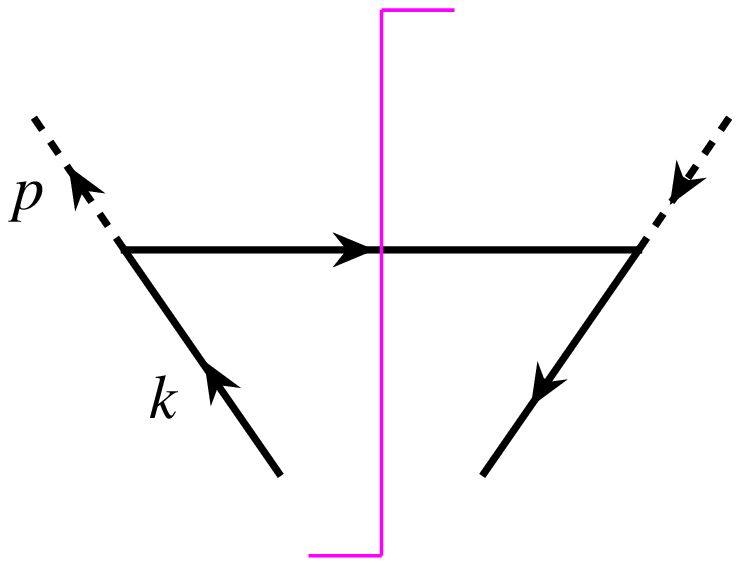, width=1.3in}
\caption{Lowest-order Feynman diagram for a quark with momentum $k$ fragmenting into a $\pi$ meson with momentum $p$.}
\label{frag}
\eef

At tree level, the fragmentation of a quark is modeled through the process $q^*\to \pi q$,
see Fig.~\ref{frag}. One can obtain the unpolarized quark fragmentation function $D(z, p_\perp^2)$ from the definition in Eq.~(\ref{main}) which has been done in \cite{Bacchetta:2002tk}. The Collins function can be calculated similarly, though one needs to go beyond the tree diagram and consider the $\pi$-loop to obtain the final result, see Ref.~\cite{Bacchetta:2002tk}.

Since the chiral quark model Lagarangian in Eq.~(\ref{lqqpi}) conserves parity, it does not generate  $\p$-odd fragmentation functions $\widetilde{D}(z, p_\perp^2)$ and $\widetilde{H}_1^\perp(z, p_\perp^2)$. As we stated in the Introduction, QCD contains topological gauge field configurations, and their effect can be mimicked by an effective space-time dependent $\theta$ field \cite{Kharzeev:1998kz, Kharzeev:2004ey}. One can thus add to the Lagrangian of QCD the term $(g^2/32\pi^2)\theta(x, t)F_a^{\mu\nu}\tilde{F}^a_{\mu\nu}$; performing an axial $U(1)$ rotation this term can be transformed into $\frac{1}{2N_f}\partial_\mu \theta \bar{\psi}_q\gamma^\mu\gamma^5\psi_q$
\cite{Fukushima:2008xe}.
Let us define an effective $\bar{\theta}_\mu=\partial_\mu \theta/2N_f$, whose zero (time) component is the chiral chemical potential $\mu_5$ introduced in \cite{Fukushima:2008xe}. The existence of this new term will yield a modified 
quark propagator 
$i\widetilde{S}(p,\bar{\theta})=i/(\sla{p}-m+\sla{\bar{\theta}}\gamma^5)$ given by 
\ben
i\tilde{S}(p, \bar{\theta})&=&i\left[\mathcal{P_R}S(p+\bar{\theta})+\mathcal{P_L}S(p-\bar{\theta})\right]
\nnu
&&
\times\left[1+m\gamma^5 \left(S(p+\bar{\theta})-S(p-\bar{\theta})\right)\right]
\nnu
&&\times\left[1+\frac{4m^2\bar{\theta}^2}{\left((p+\bar{\theta})^2-m^2\right)\left((p-\bar{\theta})^2-m^2\right)}\right]^{-1}
\label{prop}
\een
where $\mathcal{P_{L,R}}$ are the left (right) projection operator $\mathcal{P_{L,R}}=(1\pm\gamma^5)/2$, 
and $iS(p)=i(\sla{p}+m)/(p^2-m^2)$ is the conventional quark propagator.

With the modified quark propagator in the $\p$-odd "bubble", both the $\p$-odd fragmentation functions can be generated at the tree level as shown in Fig.~\ref{frag}.
We further find that the new fragmentation functions are suppressed by a factor $\bar{\theta}^{0,3}/p^+$ when $\bar{\theta}$ is along time or $z$-direction. 
Since $p^+$ is a large component in our twist-expansion, we thus neglect the effect from the $0,3$ components of $\bar \theta$ to be self-consistent. On the other hand, if the $\bar\theta$ is along the transverse direction that is perpendicular to $p_\perp$, we find that the $\p$-odd fragmentation functions vanish. We thus only consider the situation when $\bar\theta$ is along the $p_\perp$ direction: $\bar\theta^\mu=\bar\theta_\perp \hat{p}_\perp^\mu$, in which case we find
\ben
\widetilde{D}(z,p_\perp^2)&=&\frac{g_A^2}{64f_\pi^2 \pi^3z}
\frac{4\bar{\theta}_\perp p_\perp}{p_\perp^2+z^2 m_q^2+(1-z)m_\pi^2}
\bigg[1-\frac{z}{2}
\nnu
&&
\left.
-\frac{4(1-z)^2 z^2 m_q^2\,m_\pi^2}{\left(p_\perp^2+z^2 m_q^2+(1-z)m_\pi^2\right)^2}
\right],
\\
%%%%%%%%%%
\widetilde{H}_1^\perp(z, p_\perp^2)&=&
\frac{g_A^2}{4f_\pi^2}\frac{m_q m_\pi}{8\pi^3 }
\frac{\bar{\theta}_\perp}{p_\perp} 
\frac{1}{\left(p_\perp^2+z^2 m_q^2+(1-z)m_\pi^2\right)^3}
\nnu
&&
\times
\bigg[\left(p_\perp^2+z^2 m_q^2\right)^2 (z-2)+
(1-z)^2 m_\pi^2
\nnu
&&
\times
\left[(3 z-2) m_\pi^2-4 (p_\perp^2-z^2 m_q^2)\right]\bigg],
\een
where we have neglected terms $\sim\mathcal{O}(\bar{\theta}^2)$. From above equations, we see that 
both of the $\p$-odd fragmentation functions are proportional to $\bar\theta_\perp/p_\perp$. Since $p_\perp$ is a small component, the effect is not suppressed. We will now estimate the size of the observable effect generated by the $\p$-odd fragmentation functions within the same model.

{\it 3.~Observable effect of parity-odd fragmentation functions.}
\label{signal}
Let us now discuss the experimental consequences of the $\p$-odd fragmentation functions. As a first step, we study a relatively simple process, the back-to-back dihadron production in $e^+e^-$ collisions $e^+e^-\to h_1h_2+X$. The method we presented here can be generalized to study $\p$-odd effects in heavy ion collisions. 

At leading order in QCD coupling, the two hadrons $h_1$ and $h_2$ in $e^+e^-$ collisions are the fragmentation products of a quark and an antiquark
originating from $e^+e^-\to q\bar q$ annihilation. Following Ref.~\cite{Anselmino:2007fs}, we choose a reference frame such that the $e^+e^-\to q\bar q$ annihilation occurs in the $x$-$z$ plane, with the back-to-back quark and antiquark moving along the $z$-axis. The final hadrons $h_1$ and $h_2$ carry light-cone momentum fractions $z_1$ and $z_2$ and have intrinsic transverse momenta $p_{1\perp}$ and $p_{2\perp}$ with respect to the directions of the fragmenting quarks. Using the fragmentation parameterization in Eq.~(\ref{main}), one can derive the differential cross section as
\ben
\frac{d\sigma}{d\mathcal{PS}}
&=&
\sigma_0\sum_q e_q^2 \bigg\{
(1+\cos^2\theta)
\nnu
&&
\times
\left[ D_q(z_1) D_{\bar{q}}(z_2) - \widetilde{D}_q(z_1) \widetilde{D}_{\bar{q}}(z_2)\right]
\nnu
&&
+\sin^2\theta \cos(\phi_1+\phi_2)
\nnu
&&\times
\left[H_q^\perp(z_1)H_{\bar{q}}^\perp(z_2)+\widetilde{H}_q^\perp(z_1)\widetilde{H}_{\bar{q}}^\perp(z_2)\right]
\nnu
&&
+\sin^2\theta \sin(\phi_1+\phi_2)
\nnu
&&\times
\left[H_q^\perp(z_1)\widetilde{H}_{\bar{q}}^\perp(z_2)-\widetilde{H}_q^\perp(z_1)H_{\bar{q}}^\perp(z_2)\right]\bigg\},
\label{eemain}
\een
where the phase space $d\mathcal{PS}=dz_1 dz_2 d\cos\theta d(\phi_1+\phi_2)$, $\sigma_0=N_c \alpha_{em}^2/4Q^2$, and
$\theta$ is the angle between the initial beam direction and the $z$-axis, not to be confused with the $\theta(x)$ field. In Eq.~(\ref{eemain}), we have integrated over the moduli of the intrinsic momenta, $p_{1\perp}$ and $p_{2\perp}$, and over the azimuthal angle $\phi_1$. The $p_\perp$-integrated functions
$D_q(z)$ and $H_q^\perp(z)$ are defined as
\ben
D_q(z)&=&\int d^2p_\perp D_q(z, p_\perp^2),
\\
H_q^\perp(z)&=&\int d^2p_\perp \frac{|\vec{p}_\perp|}{M}H_1^{\perp q}(z, p_\perp^2).
\een
The definition of $\widetilde{D}_q(z)$ (or $\widetilde{H}_{\bar{q}}^\perp(z)$) is similar to $D_q(z)$ (or $H_q^\perp(z)$).

The $\cos(\phi_1+\phi_2)$ correlation is usually referred to as the Collins effect, analyzed recently by BELLE Collaboration \cite{Seidl:2008xc}. However, we find that the product of two $\p$-odd fragmentation functions $\widetilde{H}_q^\perp(z)$ leads to the same azimuthal correlation. This complicates the extraction of the Collins function, but may in effect provide an alternative view of the origin of the Collins effect and puts an experimental  constraint of the $\p$-odd fragmentation function $\widetilde{H}_q^\perp(z)$. It is interesting that a new azimuthal correlation also emerges: the $\sin(\phi_1+\phi_2)$ term, which is explicitly $\p$-odd. Notice that for the $\sin(\phi_1+\phi_2)$ contribution, the first term corresponds to the situation when the antiquark fragments inside the $\p$-odd bubble, see Fig.~\ref{qqbar}(a), whereas the second term corresponds to the situation when the quark fragments inside the $\p$-odd bubble, see Fig.~\ref{qqbar}(b). They have the opposite sign, and thus when averaged over the large number of events, the effect will vanish. Thus a $\p$-odd effect happens only on the event-by-event basis \cite{Kharzeev:2007jp}. 
\bef
\psfig{file=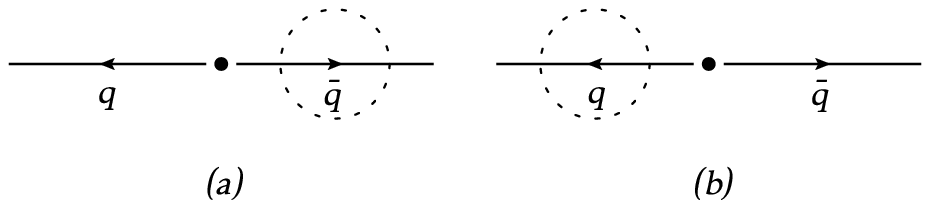, width=0.45\textwidth}
\caption{Illustration of the antiquark (a) and the quark (b) fragmenting in the $\p$-odd bubble.}
\label{qqbar}
\eef

To estimate the effect, let us assume that the antiquark fragments inside the $\p$-odd bubble; the relative magnitude of the correlation will depend on the following factor $I(\bar{\theta}, z_1, z_2)$, besides the kinematic factor $\sin^2\theta/(1+\cos^2\theta)$,
\ben
I(\bar{\theta}, z_1, z_2)=\frac{H_q^\perp(z_1)\widetilde{H}_{\bar{q}}^\perp(z_2)}{D_q(z_1) D_{\bar{q}}(z_2) - \widetilde{D}_q(z_1) \widetilde{D}_{\bar{q}}(z_2)}.
\een
Certainly $I(\bar{\theta}, z_1, z_2)$ depends on the size of $\bar{\theta}_\perp$. To estimate $\bar{\theta}_\perp$, we resort to the instanton vacuum model (for a review, see \cite{Schafer:1996wv}). According to \cite{Schafer:1996wv}, the two most important parameters are the mean size of the instanton $\rho\sim 1/3$ fm and the typical separation $R$ between instantons which satisfies $\rho/R\sim 1/3$. We thus estimate $\bar \theta_\perp$ as follows:
\ben
\bar \theta_\perp\sim \frac{1}{2N_f}\partial_\perp \theta(\vec{x}, t)\sim \frac{1}{2N_f}\cdot\frac{1}{\rho}\cdot\frac{\rho^2}{R^2}\sim10~{\rm MeV},
\een
where we have used $N_f=3$, and the factor $\rho^2/R^2$ represents the probability for a quark scatters off an instanton; note that in Minkowski space-time the instanton event is elongated along the light cone \cite{Efremov:1995ff}.
With $\bar \theta_\perp=10$ MeV, and other standard parameters of the chiral quark model \cite{Manohar:1983md}, and using the calculation of the fragmentation functions taken from Ref.~\cite{Bacchetta:2002tk}, we find $I(\bar{\theta}, z_1, z_2)\sim1.5\%$
for a typical $z_1=z_2=0.5$ at BELLE experiment,
with the final two hadrons as $\pi^+$ and $\pi^-$. We urge the experimentalists at BELLE, RHIC and elsewhere to carry out an analysis to constrain the $\p$-odd fragmentation functions. Because of the universality of the fragmentation functions, we expect that the formalism developed here could be generalized to other processes.
\vskip0.2cm
{\it 4.~Conclusion.}
\label{sum}
In this letter we have studied the quark fragmentation in the topologically non-trivial QCD background. We have found two new fragmentation functions besides the well-known spin-averaged fragmentation function and the Collins function. Both of the new fragmentation functions are $\p$-odd. We have related the magnitude of these functions to the typical size of the topological fluctuations (described by the effective $\theta(x)$ field). We have studied the 
observable effects of the $\p$-odd fragmentation functions in back-to-back dihadron production in $e^+e^-$ collisions, and have found that a new azimuthal correlation $\propto \sin(\phi_1+\phi_2)$ appears. Since the new azimuthal correlation is explicitly $\p$-odd, it can be
observed only on an event-by-event basis. Our results also offer a new interpretation of the Collins correlation. We encourage the experimentalists to carry out  an analysis to constrain the $\p$-odd fragmentation functions, and anticipate new applications, e.g. in heavy ion collisions.
\vskip0.2cm
We thank J.~Liao, R.~Millo, M.~Grosse~Perdekamp, J.~Qiu, E.~Shuryak, S.~Taneja, A.~Vossen and F.~Yuan for helpful discussions and useful comments. We are grateful to RIKEN, Brookhaven National Laboratory, 
and the U.S. Department of Energy 
(Contract No.~DE-AC02-98CH10886) for supporting this work.

%%%%%%%%%%%%%%

\end{document}